\documentclass[aps, prl, twocolumn,superscriptaddress,nofootinbib,10pt]{revtex4-1}
\usepackage{graphicx}
\usepackage{dcolumn}
\usepackage{amssymb,amsmath,amsthm,mathrsfs}
\usepackage{epstopdf}
\usepackage{lipsum}
\usepackage{hyperref}
\usepackage{empheq}
\usepackage{color}
\usepackage[normalem]{ulem} 
\usepackage{physics}
\usepackage[normalem]{ulem}
\usepackage[ruled,vlined]{algorithm2e}

\newcommand{\be}{\begin{equation}}
\newcommand{\ee}{\end{equation}}
 \newcommand{\bea}{\begin{eqnarray}}
\newcommand{\eea}{\end{eqnarray}}

\newcommand{\re}[1]{(\ref{#1})}

\renewcommand{\d}{\delta}

\newcommand{\rmd}{\mathrm{d}}

\newcommand{\de}{\mathrm{d}}

\newcommand{\mr}{\mathrm}

\newcommand{\phil}{{\bar\phi}}
\newcommand{\phif}{{\bar\phi}_\mr{f}}
\newcommand{\phii}{{\bar\phi}_\mr{i}}
\newcommand{\phipbh}{{\bar\phi}_\mr{PBH}}
\newcommand{\pil}{{\bar \pi}}

\newcommand{\R}{\mathcal{R}}
\newcommand{\PR}{\mathcal{P}_\mathcal{R}}
\newcommand{\kc}{k_\mr{c}}
\newcommand{\kpbh}{k_\mr{PBH}}

\begin{document}

\rightline{HIP-2020-32/TH}
\vspace*{0.5cm}

\title{Non-Gaussian tail of the curvature perturbation in stochastic ultra-slow-roll inflation: implications for primordial black hole production}

\newcommand{\addressIFIC}{Instituto de F\'isica Corpuscular (IFIC), CSIC-Universitat de Valencia, Spain}
\newcommand{\addressHU}{University of Helsinki, Department of Physics and Helsinki Institute of Physics, P.O. Box 64, FIN-00014 University of Helsinki, Finland}
\newcommand{\addressTallinn}{Laboratory of High Energy and Computational Physics, National Institute of Chemical
Physics and Biophysics, R\"{a}vala pst. 10, 10143 Tallinn, Estonia}

\author{Daniel G. Figueroa} \affiliation{\addressIFIC}
\author{Sami Raatikainen} \affiliation{\addressHU}
\author{Syksy R\"{a}s\"{a}nen} \affiliation{\addressHU}
\author{Eemeli Tomberg} \affiliation{\addressTallinn}

\date{\today}

\begin{abstract}
We consider quantum diffusion in ultra-slow-roll (USR) inflation. Using the $\Delta N$ formalism, we present the first stochastic calculation of the probability distribution $P(\R)$ of the curvature perturbation during USR. We capture the non-linearity of the system, solving the coupled evolution of the coarse-grained background with random kicks from the short wavelength modes, simultaneously with the mode evolution around the stochastic background. This leads to a non-Markovian process from which we determine the highly non-Gaussian tail of $P(\R)$. Studying the production of primordial black holes in a viable model, we find that stochastic effects during USR increase their abundance by a factor $\sim 10^5$ compared to the Gaussian approximation.
\end{abstract}


\maketitle

{\it Introduction.--}
Compelling evidence~\cite{Akrami:2018odb} supports a phase of accelerated expansion, inflation, as the leading framework for the early universe~\cite{Starobinsky:1980te, Kazanas:1980tx, Guth:1980zm, Sato:1980yn, Mukhanov:1981xt, Linde:1981mu, Albrecht:1982wi, Hawking:1981fz, Chibisov:1982nx, Hawking:1982cz, Guth:1982ec, Starobinsky:1982ee, Sasaki:1986hm, Mukhanov:1988jd}. In the simplest models, a scalar field -- the inflaton -- rolls down its potential with the Hubble friction and potential push balanced. This is known as slow-roll (SR). However, if the potential has a very flat section or a shallow minimum, the potential push becomes negligible, and the inflaton velocity falls rapidly due to Hubble friction. This is called ultra-slow-roll (USR) \cite{Faraoni:2000vg, Kinney:2005vj, Martin:2012pe, Dimopoulos:2017ged, Pattison:2018}. 
While SR generates close to scale-invariant and almost Gaussian perturbations, as observed in the cosmic microwave background (CMB), the perturbations produced by USR are far from scale-invariant and can be highly non-Gaussian.
This implies that the inflaton cannot be in USR when the observed CMB perturbations are generated.\footnote{As studied in \cite{Ando:2020fjm}, USR between the observable CMB region and the end of inflation can nevertheless affect the CMB spectrum through rare patches that undergo a large number of e-folds of inflation. The effect of USR near the CMB region on the power spectrum has also been studied in \cite{Enckell:2020lvn}.}
However, if the inflaton enters USR afterwards, large perturbations can be created on small scales, potentially seeding primordial black holes (PBH)~\cite{Garcia-Bellido:2017mdw, Ezquiaga:2017fvi, Kannike:2017bxn, Hertzberg:2017dkh, Germani:2017bcs, Motohashi:2017kbs, Gong:2017qlj, Ballesteros:2017fsr, Rasanen:2018fom, Drees:2019xpp, Biagetti:2018pjj, Ezquiaga:2018gbw, Ballesteros:2020sre, Fu:2020lob}, a longstanding dark matter candidate~\cite{Chapline:1975ojl, Dolgov:1992pu, Jedamzik:1996mr, Ivanov:1994pa, GarciaBellido:1996qt, Yokoyama:1995ex, Ivanov:1997ia, Blais:2002nd}.

During inflation, initially sub-Hubble ($k \gg aH$) quantum fluctuations are amplified and stretched to super-Hubble scales ($k \ll aH$), where $k$ is the comoving wavenumber, $a$ is the scale factor and $H\equiv\dot{a}/a$ is the Hubble rate. Once modes reach super-Hubble scales, they can be coarse-grained, contributing stochastic noise to the evolution of the background formed by long wavelength modes, which are squeezed and 'classicalized' \cite{Calzetta:1995ys,Polarski:1995jg,Lesgourgues:1996jc,Kiefer:1998jk,Kiefer:1998qe,Kiefer:2008ku}.
This is described by the formalism of stochastic inflation \cite{Starobinsky:1986fx, Starobinsky:1994bd, Morikawa:1989xz, Habib:1992ci, Mijic:1994vv, Bellini:1996uh, Martin:2005hb, Tsamis:2005hd, Woodard:2005cv, vanderMeulen:2007ah, Finelli:2008zg, Beneke:2012kn, Gautier:2013aoa, Garbrecht:2013, Levasseur:2013ffa, Levasseur:2013tja, Levasseur:2014ska, Garbrecht:2014dca, Burgess:2014eoa, Burgess:2015ajz, Onemli:2015, Boyanovsky:2015tba, Boyanovsky:2015jen, Moss:2016uix, Collins:2017haz, Prokopec:2017vxx, Tokuda:2017fdh, Tokuda:2018eqs, Pinol:2018euk, Grain:2017dqa, Vennin:2015hra, Pattison:2017mbe, Pinol:2020cdp, Ando:2020fjm}. Stochastic effects can be particularly relevant during USR for two reasons: $i)$ the classical push from the potential is negligible, so the inflaton velocity decays rapidly and the background evolution is more sensitive to stochastic kicks, $ii)$ the perturbations are larger and hence give stronger kicks~\cite{Vennin:2015hra, Biagetti:2018pjj, Ezquiaga:2018gbw, Cruces:2018cvq, Firouzjahi:2018vet, Pattison:2019hef, Ballesteros:2020sre, De:2020hdo, Ando:2020fjm, Vennin:2015hra, Firouzjahi:2018vet, Pattison:2017mbe}.

Stochastic effects on the power spectrum ${\mathcal P}_{\R}(k)$ of the curvature perturbation $\R$ generated during USR have been studied in~\cite{Vennin:2015hra, Biagetti:2018pjj, Ezquiaga:2018gbw, Cruces:2018cvq, Firouzjahi:2018vet, Pattison:2019hef, Ballesteros:2020sre, De:2020hdo, Ando:2020fjm} (see~\cite{Vennin:2015hra, Firouzjahi:2018vet, Ezquiaga:2018gbw} for higher moments). It was demonstrated in \cite{Pattison:2017mbe, Ezquiaga:2019ftu}, however, that stochastic effects lead to an exponential tail in the probability distribution $P(\R)$, which overtakes the linear theory Gaussian tail. Calculating the power spectrum ${\mathcal P}_{\R}(k)$ is therefore not enough to determine the PBH abundance today, $\Omega_{\rm PBH}$, which is exponentially sensitive to the shape of the tail of $P(\R)$. In this Letter we present the first calculation of the non-Gaussian tail of $P(\R)$ due to stochastic effects during USR. We solve simultaneously the evolution of the background dynamics with stochastic kicks from the small wavelength modes, and the evolution of the small wavelength modes that live in this stochastic background. We consider a model where the Standard Model Higgs is the inflaton~\cite{Bezrukov:2007ep, Rubio:2018ogq}. We use the renormalization group running to create a shallow minimum that leads to USR, tuned to produce PBHs with mass $M_{\rm PBH}=7\times10^{-15}M_{\odot}$, with an abundance that contributes significantly to dark matter in the Gaussian approximation~\cite{Rasanen:2018fom} (see also~\cite{Ezquiaga:2017fvi,Ballesteros:2017fsr,Kannike:2017bxn, Ballesteros:2017fsr, Bezrukov:2017dyv, Hertzberg:2017dkh}). We adjust the SR part of the potential by hand to fit CMB observations.

{\it Stochastic formalism.--} We consider a spatially flat Friedmann--Lema\^itre--Robertson--Walker (FLRW) background metric with scalar perturbations, split into long and short wavelength modes. Correspondingly, the inflaton is decomposed as
${\phi = \phil(t,\vec{x}) + \delta\phi(t,\vec{x})}$,
where
$\phil = (2\pi)^{-3/2}\int_{k < \kc} \rmd^3 k \, \phi_{\vec{k}}(t) e^{-i\vec{k} \cdot \vec{x}}$
and
$\delta\phi = (2\pi)^{-3/2}\int_{k > \kc} \rmd^3 k \, \phi_{\vec{k}}(t) e^{-i\vec{k} \cdot \vec{x}}$. The long wavelength part $\phil$ describes the inflaton coarse-grained over a super-Hubble patch of length $2\pi/\kc$, where $\kc = \sigma aH$ is a coarse-graining scale with $\sigma\ll1$ (we discuss the precise value later).

In the leading long wavelength approximation, the background follows the Friedmann equations, while the short wavelength modes obey the linear perturbation equations over the FLRW background  \cite{Salopek:1992qy, Wands:2000dp}. As the universe expands, short wavelength modes are stretched to super-Hubble scales. Going beyond the leading approximation, the resulting change in the local background is captured by the stochastic formalism, where the background evolution is given by a Langevin equation that includes the backreaction of the short wavelength perturbations \cite{Starobinsky:1986fx, Starobinsky:1994bd, Morikawa:1989xz, Habib:1992ci, Mijic:1994vv, Bellini:1996uh, Martin:2005hb, Tsamis:2005hd, Woodard:2005cv, vanderMeulen:2007ah, Finelli:2008zg, Beneke:2012kn, Gautier:2013aoa, Garbrecht:2013, Levasseur:2013ffa, Levasseur:2013tja, Levasseur:2014ska, Garbrecht:2014dca, Burgess:2014eoa, Burgess:2015ajz, Onemli:2015, Boyanovsky:2015tba, Boyanovsky:2015jen, Moss:2016uix, Collins:2017haz, Prokopec:2017vxx, Tokuda:2017fdh, Tokuda:2018eqs, Pinol:2018euk, Grain:2017dqa, Vennin:2015hra, Pattison:2017mbe, Pinol:2020cdp, Ando:2020fjm}. The short wavelength modes contribute random noise to the local background equations. The randomness is due to the quantum origin of the initial conditions of the short wavelength modes.

Except for a few studies (e.g.~\cite{Levasseur:2013ffa, Levasseur:2013tja, Levasseur:2014ska}), previous works solved the short wavelength modes over a non-stochastic background. We go one step further by including the effect of the stochastic change of the local background on the dynamics of the short wavelength modes, capturing the interaction between the modes and the background at every moment. This leads to a non-Markovian process. The noise depends on the short wavelength modes, which depend on the coarse-grained field, so each new kick is affected by the history of previous kicks.

The equations of motion of the coarse-grained field with stochastic effects are obtained as usual, including the short wavelength contribution in the time derivatives only, reinterpreted as stochastic noise. For the short wavelength modes, we use linear perturbation theory in the spatially flat gauge, and replace the background fields by their coarse-grained counterparts. The equations of motion read (with the reduced Planck mass set to unity)
\begin{align}
& \phil' = \pil + \xi_\phi \label{eq:eom_phi} \, , \\
& \pil' = -( 3 + {H}'/{H} ) \pil - V_{,\phil}/H^2 + \xi_\pi \label{eq:eom_pi} \, , \\
& 2 V = ( 6 - \pi^2 ) H^2 \, , \label{eq:eom_H} \\
& \delta\phi_{\vec{k}}'' + \left(3 + {H}'/{H}\right)\hspace{-0.5mm}\delta\phi_{\vec{k}}' + \omega_k^2 \delta\phi_{\vec{k}} = 0 \, , \label{eq:eom_pert}
\end{align}
where $V(\phil)$ is the inflaton potential, $N\equiv\ln(a/a_*)$ is the number of e-folds ($*$ refers to the Hubble exit of the CMB pivot scale $k_*=0.05$ Mpc$^{-1}$), $'\equiv\rmd/\rmd N$, $\xi_\phi$ and $\xi_\pi$ are the field and momentum noise (following Gaussian statistics), respectively, and $\omega_k^2 \equiv k^2/(a H)^2 +\pil^2 (3 + H'/H) + 2\pil V_{,\phil}/H^2 + V_{,\phil\phil}/H^2$. We initialize the modes deep inside the Hubble radius in the Bunch--Davies vacuum, so $\delta\phi_{\vec{k}} = 1/(a\sqrt{2k})$, $\delta\phi_{\vec{k}}' = -(1 + i\frac{k}{aH})\delta\phi_{\vec{k}}$. We separate short and long wavelength modes with a step function in momentum space, so $\xi_\phi$ and $\xi_\pi$ are white noise, with $\langle \xi_\phi(N_1) \xi_\phi(N_2) \rangle = \frac{k^3}{2\pi^2} (1 + H'/H) |\d\phi_{\vec k}|^2|_{k=\sigma aH} \delta(N_1 - N_2)$, and analogous correlator for $\xi_\pi$ \cite{Pattison:2019hef}. The time evolution of $\phil$ receives stochastic kicks at every finite step with variance $\langle \Delta \phil^2 \rangle = \rmd N \frac{k^3}{2\pi^2} (1 + H'/H) |\d\phi_{\vec k}|^2|_{k=\sigma aH} $, where $ \rmd N$ is the time step of the numerical calculation. As the perturbations are highly squeezed (as we will discuss shortly), the momentum kicks are strongly correlated with the field kicks, $\Delta \pil = \mr{Re}(\delta\phi'_{\vec{k}}/\delta\phi_{\vec{k}}) \Delta \phil$.

\begin{figure}[t!]
\centering
\includegraphics[scale=1]{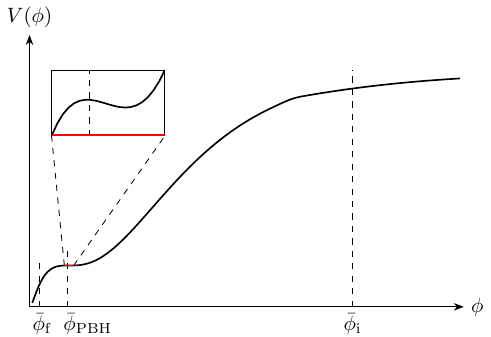}
\caption{The inflationary potential, with a plateau and a shallow local minimum. The initial field value $\phii$ (close to the CMB pivot scale), the end of USR $\phipbh$, and the end of inflation $\phif$ (where the simulation ends) are marked. The vertical axis of the inset is stretched relative to the main plot to better bring out the shape of the potential close to the minimum.}
\label{fig:V}
\end{figure}

{\it Inflation model.--}
We consider an inflaton potential $V(\phi)$ where the CMB perturbations are generated at a plateau, and there is a shallow local minimum at smaller field values, as shown in Fig.~\ref{fig:V}. The inflaton starts in SR, enters USR as it rolls over the minimum, and then returns to SR until the end of inflation. We consider a model where the Standard Model Higgs is the inflaton and the local minimum is produced by quantum corrections~\cite{Rasanen:2018fom}, tuned to produce PBHs with mass $M_{\rm PBH}=7\times10^{-15}M_\odot$, with an abundance that roughly agrees with the observed dark matter density in the Gaussian approximation. Contrary to~\cite{Rasanen:2018fom}, here we adjust the plateau by hand to fit CMB observations~\cite{Akrami:2018odb}. We give the details in the Supplemental Material. At the CMB pivot scale $k_*$ the spectral index is $n_s=0.966$ and the tensor-to-scalar ratio is $r=0.012$. 

{\it Squeezing and classicalization.--}
For the stochastic formalism to be valid, the perturbations must be classical by the time they join the background. Classicality can be characterized by squeezing of the mode wave functions. A squeezed state can be written as~\cite{Polarski:1995jg,Grain:2019vnq}
\begin{eqnarray}
    \ket{\psi} = \exp[\frac{1}{2}\qty(s^*\hat{a}^2 - s\hat{a}^\dagger{}^2)] \ket{0} \, ,
\end{eqnarray}
where $s=re^{2i\varphi}$ is the squeezing parameter, and $\hat a, \hat a^\dagger$ are standard ladder operators that satisfy $\comm{\hat{a}}{\hat{a}^ \dagger} = 1$. They determine the vacuum state, $\hat{a}\ket{0} = 0$, with respect to which the squeezing is measured. The amplitude $r$ indicates how squeezed the state is, and the phase $\varphi$ gives the squeezing direction in phase space.

Choosing $Q_{\vec{k}} = \sqrt{k}a \delta\phi_{\vec{k}}$ and $P_{\vec{k}} = a^2 H\delta\phi'_{\vec{k}}/\sqrt{k}$ for the canonical variables that define the vacuum, leads to the Bunch--Davies vacuum for the sub-Hubble modes. The corresponding operators are related to the ladder operators in the usual way, and we have
\begin{eqnarray} \label{eq:r_in_QP}
    \expval{\hat{Q}_{\vec{k}}^2 + \hat{P}_{\vec{k}}^2}{\psi_{\vec{k}}} = \cosh(2 r_k) \, .
\end{eqnarray}
The value of $r_k$ is then a proxy for classicalization. For the Bunch--Davies vacuum, the mode initially has the minimum uncertainty wave packet, for which $r_k=0$, and $r_k$ grows as the phase space probability distribution gets squeezed. Large $r_k$ implies that the probability distribution covers a large region in phase space, where the expectation value of the commutator $[\hat Q_{\vec{k}} , \hat P_{\vec{k'}}] = i\delta(\vec{k}-\vec{k'})$ is negligible compared to expectation values such as $\expval{\hat{Q}_{\vec{k}} \hat{P}_{\vec{k'}}+ \hat{P}_{\vec{k}}\hat{Q}_{\vec{k'}}}{\psi_{\vec{k}}}$. Thus, all relevant expectation values can be reproduced by a classical probability distribution. Squeezing makes the operators $\hat{Q}_{\vec{k}}$ and $\hat{P}_{\vec{k}}$ highly correlated, so the field and momentum kicks become approximately proportional to each other. Note that $r_k \gg 1$ corresponds to a large occupation number.

Modes get more squeezed as they are pushed further outside the Hubble radius. The coarse-graining parameter $\sigma$ has to be small enough to ensure that the mode probability distribution is sufficiently classical. However, the larger the value of $\sigma$, the more interactions between the short and long wavelength modes we capture. We choose the value $\sigma = 0.01$, for which all modes satisfy $\cosh(2r_k) > 100$ when they exit the coarse-graining scale.

{\it Gauge-dependence.--}
The perturbation equation of motion~\re{eq:eom_pert} is written in the spatially flat gauge, which is convenient for calculating the mode functions, whereas the stochastic equations \re{eq:eom_phi}, \re{eq:eom_pi} for the background are written in the uniform-$N$ gauge, as $N$ does not receive kicks. As shown in~\cite{Pattison:2019hef}, the correction to the mode functions when changing from the flat gauge to the uniform-$N$ gauge is small both in SR and USR. We have checked numerically that in our calculation this holds at all times, including during transitions between SR and USR, except for a small subset of the modes that have little quantitative impact. Gauge difference therefore has negligible impact on our results.

{\it $\Delta N$ formalism.--} We aim to calculate the coarse-grained comoving curvature perturbation $\R$ in a given patch of space, since this determines whether the patch collapses into a PBH. We use the $\Delta N$ formalism \cite{Sasaki:1995aw, Sasaki:1998ug, Wands:2000dp, Lyth:2004gb}, where $\R$ is given by the difference between the number of e-folds $N$ of the local patch and the mean number of e-folds $\bar{N}$, measured between an initial unperturbed hypersurface with fixed initial field value $\phii$, and a final hypersurface of constant field value $\phif$,
\begin{equation}
    \R = N - \bar{N} \equiv \Delta N \, .
\end{equation}
When we solve the stochastic equations, we follow a patch of size determined by the coarse-graining scale $k_c = \sigma a H$, which changes in time. The patch size at the end of the calculation gives the PBH scale we probe; we fix this to the value $\kpbh$, which we discuss below. To ensure that $\kpbh$ gives the final patch size, we stop the time evolution of $\kc$ once $\kc = \kpbh$. After this, no modes from $\delta \phi$ contribute to $\phil$: the stochastic noise is switched off, so modes with larger $k$ do not give kicks. This is a meaningful procedure as perturbations with wavelengths smaller than the size of the collapsing region should not affect PBH formation; they behave as noise that is averaged out in the coarse-graining process. We continue to evolve the local background without kicks until the end of inflation, where the field value is $\phif$. We record the final value of $N$ for each simulation, and build statistics over many runs to find the probability distribution $P(N)$. The numerical algorithm is described in the Supplemental Material.

{\it PBH production.--}
When a perturbation of wavenumber $k$ re-enters the Hubble radius during the radiation-dominated phase after inflation, it may collapse into a black hole of mass
\begin{eqnarray} \label{eq:M_PBH}
    M = \frac{4}{3}\pi \gamma H_k^{-3} \rho_k \approx 5.6 \times 10^{15} \gamma \left(\frac{k}{k_*}\right)^{-2} M_\odot \, ,
\end{eqnarray}
where $M_\odot \approx 2 \times 10^{33}$ g, $\gamma \approx 0.2$ is a parameter characterizing the collapse \cite{Carr:1975qj}, and $H_k$ and $\rho_k$ are, respectively, the background Hubble rate and energy density at Hubble entry. We assume standard expansion history.

The abundance depends on whether the collapse is computed from a peak analysis or from density threshold considerations, on how the threshold is chosen and the mapping between $\R$ and the density contrast~\cite{Carr:1975qj, Niemeyer:1999ak,Musco:2004ak,Harada:2013epa, Young:2014ana, Musco:2012au, Motohashi:2017kbs, Young:2014ana, Musco:2020jjb, Germani:2018jgr, Musco:2018rwt, Kehagias:2019eil, Young:2019yug, Escriva:2019nsa, Escriva:2020tak, Germani:2019zez, Escriva:2019phb, Young:2019osy, Wu:2020ilx, Tokeshi:2020tjq, Young:2019osy, Wu:2020ilx, Tokeshi:2020tjq}. To highlight the impact of the stochastic effects during USR versus the Gaussian approximation, we simply consider a treatment where the collapse occurs if the curvature perturbation exceeds the threshold $\R_c=1$. The fraction of simulations where $\R > \R_c$ gives the initial PBH energy density fraction $\beta$. Since PBHs behave as matter, this fraction grows during radiation domination, and today it is
\begin{equation} \label{eq:omega_PBH}
    \Omega_\text{PBH} \approx 9 \times 10^7 \gamma^{\frac{1}{2}} \beta \left(\frac{M}{M_\odot}\right)^{-\frac{1}{2}} \, .
\end{equation}
It is often assumed that $\R$ follows a Gaussian distribution (e.g.~\cite{Green:2004wb, Young:2014ana}), with variance $\sigma_\R^2 = \int_{k_\mr{IR}}^{\kpbh} \rmd (\ln k) \PR(k)$, where ${k_\mr{IR}}$ is a cutoff corresponding roughly to the size of the present Hubble radius, and whose precise value makes no difference to our results. The Gaussian approximation gives
\begin{equation} \label{eq:gaussian_beta_0}
    \beta = 2\int_{\R_c}^\infty \rmd\R \frac{1}{\sqrt{2\pi} \sigma_\R} e^{-\frac{\R^2}{2\sigma_\R^2}}
    \approx \frac{\sqrt{2}\sigma_\R}{\sqrt{\pi}\R_c}e^{-\frac{\R_c^2}{2\sigma_\R^2}} \, ,
\end{equation}
with the conventional factor of 2 \cite{Press:1973iz}. Our model is fine-tuned to give a substantial PBH abundance in the Gaussian approximation. We want to capture all the strong perturbations generated during USR, so we choose $\kpbh = e^{33.6}k_*$, which exits the Hubble radius at the end of USR, and corresponds to $M = 7 \times 10^{-15} M_\odot$. PBHs of this mass can constitute all of the dark matter \cite{Carr:2020xqk, Green:2020jor, Carr:2020gox}. In the Gaussian approximation we obtain $\sigma_\R^2=0.0149$ and $\beta = 2.7 \times 10^{-16}$. Using \eqref{eq:omega_PBH} this leads to an abundance $\Omega_\mr{PBH} = 0.13$. However, we will see below that this Gaussian approximation severely underestimates the PBH abundance.

In reality, all PBHs will not have exactly the same mass. The mass distribution could be estimated by varying $\kpbh$. However, USR produces a sharp peak in the perturbations, corresponding to a strongly peaked distribution of PBH masses. To keep the discussion simple, we stick to the value $M = 7 \times 10^{-15} M_\odot$.

\begin{figure}[t!]
\centering
\includegraphics[scale=1]{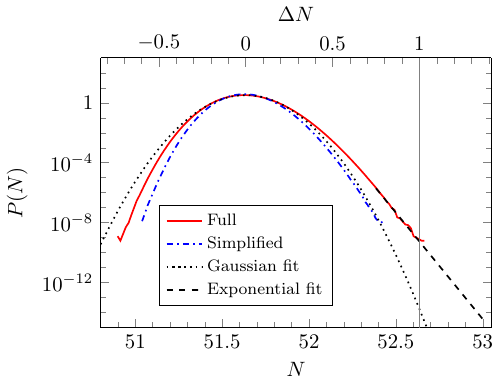}
\caption{The probability distribution for the number of e-folds. The bottom label indicates the total number of e-folds, the top label indicates deviation from the mean. The red solid line is the full result, the black dotted line is a Gaussian fit to all points, and the black dashed line is an exponential fit to the tail. The blue dash-dotted line is the simplified case discussed in the text. The vertical line marks the collapse threshold $\Delta N = 1$.}
\label{fig:P}
\end{figure}

{\it Results.--}
We have run over $10^{11}$ simulations (using over 1 million CPU hours) to find the distribution $P(N)$ of the number of e-folds between the CMB pivot scale and the end of inflation, see Fig.~\ref{fig:P}. The red solid line is the full numerical result, and the dotted black line is a Gaussian fit. The deviation from Gaussianity is evident for $|\Delta N|\gtrsim0.5$. Although the stochastic kicks push the field in either direction with equal probability, it is more likely to spend a longer (rather than shorter) time in the USR region, because the field slows down there, skewing $\Delta N$ towards positive values. The Gaussian fit has variance $\sigma_\R^2=0.0152$, close to the Gaussian estimate used to build the potential, and gives $\beta=5.3 \times 10^{-16}$.

Our data reaches $\Delta N = 1$, but the region $\Delta N >0.95$ is poorly sampled. The mean is $\bar{N}=51.64$. We estimate that resolving the tail of the distribution beyond $\Delta N = 1$ would require $10^2$ times more simulations, which translates into $\sim$100 million CPU hours. The distribution for $\Delta N\gtrsim0.6$ is well fit by a single exponential. The black dashed line in Fig.~\ref{fig:P} shows the best-fit $P(N)=e^{A-B N}$ to the data between $\Delta N=0.75$ and $\Delta N=0.95$. A jackknife analysis where we divide our data into 20 subsamples gives the mean values and error estimates $A = 1699 \pm 61$, $B = 32.7 \pm 1.2$. The mean and the best-fit are very close. To determine the PBH abundance, we extrapolate this exponential beyond the resolved region. As the abundance falls steeply, the dominant contribution comes from just beyond the threshold $\Delta N=1$. We get the PBH abundance $\beta=2\int_{\bar N+1}^{\infty}\rmd N P(N)=2B^{-1} e^{A-B (\bar N+1)} = 3.4 \times 10^{-11}$, which corresponds to $\Omega_\mr{PBH} = 1.6 \times 10^4$. The jackknife analysis gives 25\% errors on these values. The difference from the Gaussian approximation for the PBH abundance today is a factor $\sim10^5$. After our results appeared, the form of $P(N)$ with an exponential tail in stochastic USR was calculated analytically \cite{Pattison:2021oen}.

The blue dash-dotted line in Fig.~\ref{fig:P} shows a simplified treatment where the modes in the noise are fixed to their SR super-Hubble limit, $|\d\phi_{\vec k}|=H/(\sqrt{2}k^{3/2})$ with $H' \ll H$, so the noise is proportional to the Hubble parameter. (The similarity of this curve with the Gaussian fit to the full computation for $\Delta N>0$ is purely accidental.) This is a usual assumption in stochastic inflation used from the original work \cite{Starobinsky:1986fx} to the most recent studies \cite{Pattison:2021oen}. It neglects the non-linear effects we capture in our simulations, both mode evolution and the stochastic change of the coarse-grained background on the evolution of the modes, which makes the process non-Markovian. Fitting an exponential to the curve and extrapolating beyond $\Delta N = 1$, the simplified treatment underpredicts the PBH abundance by three orders of magnitude, underlining the importance of mode evolution in USR.

{\it Conclusions.--}
Applying the $\Delta N$ formalism, we find that stochastic effects in USR generate an exponential tail in the probability distribution $P(\R)$ of the curvature perturbation, as generally expected~\cite{Pattison:2017mbe, Ezquiaga:2019ftu}. Considering a realistic model, tailored to fit CMB observations and to give roughly the observed dark matter abundance in PBHs (of mass $M=7\times10^{-15} M_\odot$) in the Gaussian approximation, we find that stochastic effects during USR increase the PBH abundance today by a factor of $\sim10^5$. Our results demonstrate that when considering PBHs seeded during USR, it is crucial to calculate the shape of the tail of the probability distribution $P(\R)$, instead of simply using the power spectrum $\mathcal{P}_\R$ based on the assumption that $P(\R)$ is Gaussian. Our calculation serves as a proof of concept that the Gaussian approximation can underestimate the PBH abundance by orders of magnitude. Similar behavior is expected in any USR scenario, with the quantitative effect depending on the model. Studying USR scenarios leading to PBHs with initial mass 5 $M_\odot$, 1800 $M_\odot$ and $10^3$ kg (to produce Planck-scale relics), we find an exponential tail in all cases, and significant discrepancy in the dark matter abundance with respect to the Gaussian case. We will report the details of this elsewhere.

As a final remark, we note that our results are sensitive to the value of $\sigma$, which gives an offset between the time a mode exits the Hubble radius and the time it is coarse-grained, when it 'kicks' the local background. In SR, modes freeze to an almost scale-invariant spectrum at super-Hubble scales, so the stochastic results are insensitive to the value of $\sigma$ as long as it is sufficiently small that modes have stopped evolving~\cite{Starobinsky:1986fx} (but not too small~\cite{Habib:1992ci, Starobinsky:1994bd, Bellini:1996uh, Grain:2017dqa}). In USR this is not the case, because the near scale-invariance is lost, and super-Hubble perturbations can also evolve for longer. The usual simplified treatment where mode evolution is neglected is oblivious to this problem. The validity of the choice of $\sigma$ (more generally, the form of the stochastic equation) should be checked with a first principle derivation of the separation between system and environment in quantum field theory. While such derivations exist for stochastic SR inflation, none of the ones with explicit Langevin equations apply to USR~\cite{Morikawa:1989xz, Mijic:1994vv, Habib:1992ci, Tsamis:2005hd, Woodard:2005cv, vanderMeulen:2007ah, Finelli:2008zg, Beneke:2012kn, Gautier:2013aoa, Garbrecht:2013, Garbrecht:2014dca, Levasseur:2014ska, Burgess:2014eoa, Burgess:2015ajz, Onemli:2015, Boyanovsky:2015tba, Boyanovsky:2015jen, Vennin:2015hra, Moss:2016uix, Collins:2017haz, Prokopec:2017vxx, Tokuda:2017fdh, Tokuda:2018eqs, Pinol:2020cdp}. The dependence on $\sigma$ suggests that USR may be a more sensitive probe of decoherence and the quantum nature of inflationary perturbations than SR. \\

{We thank Nuutti Auvinen for participation in testing our code, David Alonso for advice on the jackknife analysis, and Vincent Vennin for discussions. DGF (ORCID 0000-0002-4005-8915) is supported by a Ram\'on y Cajal contract with Ref.~RYC-2017-23493,  and by the grant ``SOM: Sabor y Origen de la Materia" under no.~FPA2017-85985-P, both from Spanish Ministry MINECO. DGF also acknowledges hospitality and support from KITP in Santa Barbara, where part of this work was completed, supported by the National Science Foundation Grant No.\ NSF PHY-1748958. This work was supported by the Estonian Research Council grants PRG803, MOBTT5, and PRG1055 and by the EU through the European Regional Development Fund CoE program TK133 ``The Dark Side of the Universe''. This work used computational resources provided by the Finnish Grid and Cloud Infrastructure (urn:nbn:fi:research-infras-2016072533). The authors wish to acknowledge CSC - IT Center for Science, Finland, for computational resources.}

\bibliography{stoc.bib}

\clearpage

\section{Supplemental Material: details of the potential} \label{sec:potential_details}
Our scalar field potential comes from quantum-corrected Higgs inflation as presented in \cite{Rasanen:2018fom}, with an added modification at the CMB scales. The potential has the form
\begin{equation}
    V[\phi(h)] = \frac{\lambda_\mathrm{eff}(h)}{4}F(h)^4 \, ,
\end{equation}
where $\lambda_\mathrm{eff}(h)$ contains one-loop Coleman--Weinberg corrections with couplings running according to the one-loop beta functions of the chiral Standard Model, and we have (in units where the reduced Planck mass is unity)
\begin{equation}
    F(h) \equiv
\begin{cases}
    \frac{h}{\sqrt{1+\xi h^2}} \, , & h<h_1= 5.5h_0 \\
    0.998 \times \frac{h+4.7h_0}{\sqrt{2.50+\xi (h+4.7h_0)^2}} \, , & h > h_2 = 7h_0 \\
    \text{interpolation} \, , & h_1<h<h_2 \, ,
\end{cases}
\end{equation}
with $\xi=38.8$, and the reference scale $h_0$ is defined through $\delta\equiv\frac{1}{\xi h_0^2} = 1.48137$. This has the standard form from \cite{Rasanen:2018fom} when $h<h_1$, but is slightly modified above this to better fit the CMB observations. The interpolation between $h_1$ and $h_2$ is done with an order five polynomial, ensuring continuous $F$, $F'$, and $F''$ on both edges. In figure \ref{fig:V}, the scale $h\sim h_1\sim h_2$ corresponds to the kink in the potential below $\phii$. The canonical field $\phi$ is related to the $h$-field by
\begin{equation}
    \frac{\de \phi}{\de h} = \frac{\sqrt{1+\xi h^2 + 6\xi^2 h^2}}{1+\xi h^2} \, .
\end{equation}
The free parameters in $\lambda_\mathrm{eff}(h)$ are tuned as described in \cite{Rasanen:2018fom} so that there is a local minimum in the potential at the scale $h=h_0$, and the second slow-roll variable is $\eta\equiv\frac{\de^2 V(\phi)}{\de\phi^2}/V(\phi) = 0.41410445743324537$ at this point. For the Gaussian PBH abundance \eqref{eq:gaussian_beta_0} to stay within a $10\%$ margin of our case, $\eta$ must be fine-tuned to four decimal places. The running self-coupling is then approximately $\lambda_\mathrm{eff}(h) \approx 2.556\times10^{-6} + 1.029\times10^{-5} \ln [\sqrt{\xi}F(h)] + 2.326\times10^{-5} \ln^2 [\sqrt{\xi}F(h)]$, though we use the full expression discussed in \cite{Rasanen:2018fom}.

\section{Supplemental Material: algorithm for solving the stochastic equations}

We use an explicit Runge--Kutta method of order 4 with fixed time step $\dd{N} = 1/256$. We consider a discrete grid of modes with modulus evenly distributed on a logarithmic scale as $\ln(k_{i+1}) = \ln(k_i) + 1/32$, for a total of about 1100 modes. The evolution of each mode begins when $k = \alpha a H$, with $\alpha = 100$ (the results are insensitive to making $\alpha$ larger). The longest wavelength mode we consider corresponds to the CMB pivot scale $k_*$, and its evolution starts immediately at the onset of each simulation. The next mode to begin evolving is denoted by $k_\mr{next}$. For each realization, the code executes the algorithm below:

\begin{algorithm}[H]
\SetAlgoVlined
\DontPrintSemicolon
\SetInd{0.5em}{1em}
\SetAlgoHangIndent{1em}
\SetVlineSkip{0.5em}
Set initial values for $N$, $\phil$, $\pil$. Set $k_\mr{next} = k_\mr{*}$. Set current kick coefficient to zero.\;
\While{$\phil > \phif$}
{
    Evolve $N$, $\phil$, $\pil$.\;
    \For{all modes k in the simulation}
    {
        \eIf{$k > \sigma a H$}
        {
            Evolve $\delta\phi_{\vec{k}}$, $\delta\phi_{\vec{k}}'$.
        }
        {
            Evolve $\delta\phi_{\vec{k}}$, $\delta\phi_{\vec{k}}'$ to $k = \sigma a H$. Set the current kick coefficient from $\delta\phi_{\vec{k}}$, $\delta\phi_{\vec{k}}'$. Remove mode $k$ from the simulation.
        }
    }
    \eIf{$k_\mr{next} \leq \kpbh$}
    {
        \If{$k_\mr{next} \leq \alpha a H$}
        {
            Add mode $k = k_\mr{next}$ to the simulation. Set initial values for $\delta\phi_{\vec{k}}$, $\delta\phi_{\vec{k}}'$. Evolve $\delta\phi_{\vec{k}}$, $\delta\phi_{\vec{k}}'$ from $k = \alpha a H$. Set $k_\mr{next} = e^{1/32} k_\mr{next}$.
        }  
    }
    {
        \If{$k_\mr{next} \leq \sigma a H$}
        {
            Set the current kick coefficient to zero.
        }   
    }

    Add stochastic kick to $\phil$, $\pil$ using the current kick coefficient.
}
\caption{Evolution for each run}
\end{algorithm} 

\end{document}